\documentclass[journal,10pt]{IEEEtran}
\usepackage{mathtools}
\usepackage{amsmath,color,amsthm}
\usepackage{breqn}
\usepackage{cite}
\usepackage{algorithm,algpseudocode}
\usepackage{graphicx}
\usepackage{caption}
\usepackage{lipsum}
\usepackage{cuted}
\usepackage{type1cm}
\usepackage{lettrine}
\usepackage{enumerate}
\usepackage{graphicx}
\usepackage{float}
\usepackage{mathrsfs}

\usepackage{amsmath,amsthm,amssymb}

\usepackage{wasysym} 

\DeclarePairedDelimiterX\Basics[1](){ #1}

\date{\today}

\begin{document}
	\title{Delta-OMA (D-OMA): A New Method for Massive Multiple Access in 6G} 

\author{Yasser Al-Eryani and Ekram Hossain\thanks{The authors are with the Department of Electrical and Computer Engineering at the University of Manitoba, Canada (Emails: \{Yasser.Aleryani, Ekram.Hossain\}@umanitoba.ca).}}

 	\maketitle
 	\begin{abstract}
 	    A new multiple access method, namely, delta-orthogonal multiple access (D-OMA) is introduced for massive access in future generation 6G cellular networks. D-OMA is based on the concept of distributed large coordinated multipoint transmission-enabled  non-orthogonal multiple access (NOMA) using partially overlapping sub-bands for NOMA clusters. Performance of this scheme is demonstrated in terms of outage capacity for different degrees of overlapping of NOMA sub-bands. D-OMA can also be used for enhanced security provisioning in both uplink and downlink wireless access networks. Practical implementation issues and open challenges for optimizing D-OMA are also discussed. 
 	\end{abstract}
	
 	\section{Introduction}

Each generation of the cellular wireless systems has been characterized by a new multiple access method.     
Specifically, the first generation (1G) systems were based on frequency division multiple access (FDMA), while the second, third and fourth generations were based on time division multiple access (TDMA), code division multiple access (CDMA) and orthogonal frequency division multiple access (OFDMA), respectively. As for the 5th generation (5G) cellular,  although many development and standardization efforts are still in progress, it is clear that there will be no revolutionary multiple access technology except the use of extremely wider range of spectrum (up to 60 GHz) and the adoption of non-orthogonal multiple access schemes (NOMA) in addition to orthogonal frequency division multiple access (OFDMA)  \cite{6G8412482,8085125,Tabassum2018}.
 The adoption of higher frequency spectrum bands in 5G air interface such as the millimeter wave (mmWave) bands would create serious propagation issues due to high path-loss and beam directivity requirements. 
 This can be partly mitigated  by ultra-dense deployment of access points (APs) which in turn requires sophisticated coordination and cooperation among distributed APs to minimize the effect of co-channel interference resulting from overlapping coverage areas of nearby cells. 

 Nevertheless, 5G is expected to provide three major unique services, namely, enhanced mobile broadband communication (eMBB), ultra-reliable low-latency communication, and massive machine type communication (mMTC) \cite{5G8540332}. 
 The objective  of eMBB is to provide  operating modes with higher data rates and extended coverage area (compared to LTE), while ultra-reliable low-latency services will provide authenticated services for critical applications such as autonomous driving and health monitoring devices.
 The role of mMTC is to control data flow to/from massive number of wireless devices with guaranteed performance level.
 
 While 5G cellular networks will embrace many distinguishing enhancements over 4G networks to provide increased transmission rates with decreased latency, enhanced system reliability and performance, decreased sizes of terminal devices, and energy-efficient hardware and network designs, the emergence of advanced technologies will drive its evolution further toward beyond 5G (B5G) or so called sixth generation (6G) cellular networks. The key drivers for the 6G cellular networks can be summarized as follows: 
 
  \begin{enumerate}[\huge .]
  \item {\em All-connected networks}: With the proliferation of internet of things (IoT) and mMTC services, every wireless  device will be connected to one or more wireless access networks to be served by multiple access points (APs) or base stations (BSs), which in turn will be connected to a general cloud network to access cloud-based services (e.g. edge-computing and caching services).  
  Examples of these applications/services include virtual reality, autonomous driving, smart city and smart grid applications, industrial control and smart manufacturing, surveillance and safety, as well as numerous health monitoring services.  The wireless devices will also have peer-to-peer connectivity through single or multi-hop  communications. Also, the terrestrial cellular systems will be integrated with the airborne (or non-terrestrial/aerial/drone-assisted) networks with mobile BSs/APs. Accordingly, the conventional cellular system models will not be sufficient to describe these new systems.  Also, these networks will be application and content-driven networks rather than only data transmission networks. Therefore, new techniques in terms of network planing and optimization will be required.
    
\item {\em Energy minimization at the device and network levels}:  Since the users, machines, APs/BSs, as well as other network nodes will need to use advanced signal processing techniques as well as process more data (e.g. for artificial intelligence-enabled applications and services) power consumption will increase significantly. Also, power consumption in the radio transceivers (e.g. at the power amplifiers, A/D and D/A converters) will need to be minimized at millimeter-wave and nanometer-wave frequencies. With ultra-dense deployment of APs as well as large deployment of edge computing/caching servers in the wireless access network, it will create a dire need for new concepts for energy conserving, energy charging, harvesting, and energy cooperation among network nodes.

 \item {\em Efficient spectrum utilization and/or more spectrum}: 5G new radio (NR) extends the frequency range of 4G networks (0.6$-$6 GHz) into some higher bands of frequencies (30$-$300 GHz band millimeter waves [mmWave] and free-space optical [FSO] at the range of 200$-$385 THz). New technologies will need to be developed for wireless access and backhauling as well as coexistence (in case of unlicensed spectrum) in these new bands. 
 
    
 \end{enumerate}

 \begin{figure*}
    \centering
    \includegraphics[width=6in]{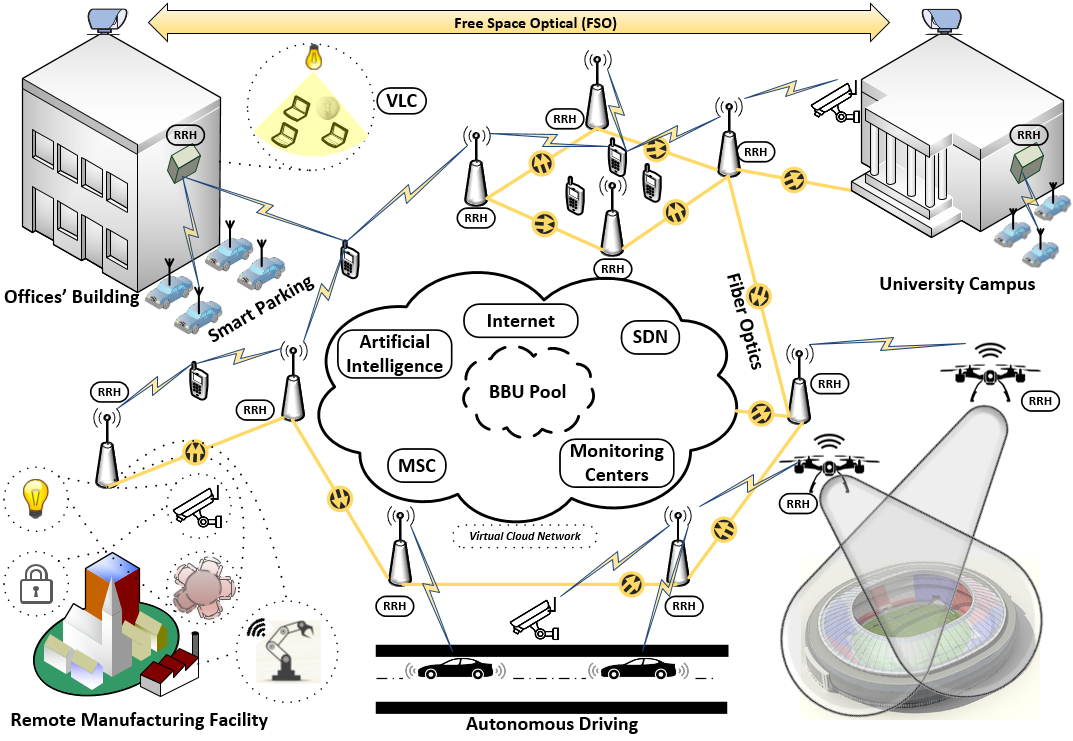}
    \caption{\textsf{Cell-less 6G network architecture.}}
    \label{6GNetwork}
\end{figure*}
\section{Cell-less Architecture for Future Wireless Networks}

Generally speaking, the concept of cell-based network architecture will not be suitable for future wireless networks, especially in urban ultra-dense wireless access scenarios in which numerous wireless devices are served simultaneously using multipoint transmissions and multipoint user associations (Fig.~\ref{6GNetwork}). Utilizing the very fast backhauling links among different BSs/APs, the overall network will appear as a cell-less distributed massive multiple-input multiple-output (massive MIMO) system from end-device's point of view. Specifically, all APs will be aware of all active devices within their vicinity. APs may be considered as remote radio heads (RRHs) as in the case of cloud radio access networks (CRANs) \cite{CRAN}. 
Every device may be served by more than on RRH either by transmission coordination or transmission multiplexing.  It may be useful to view this cell-less architecture as a generalized version of the well known coordinated multipoint transmission (CoMP) at which cooperating APs jointly serve all devices within their coverage area (cell-edge and cell-centre devices).
This can be enabled by the utilization of very fast centralized processing units that assign resources to different terminal devices while data processing may be conducted at the so-called baseband unit pool (BBU) as in the case of CRANs. With complete coordination among different RRHs, interference management
can be performed optimally or near optimally throughout some centralized or distributed optimization methodologies.

Such a network architecture will need to connect millions of devices (e.g. mMTC devices) for which automated services are required to be provided without direct human interactions. Traditional orthogonal multiple access (OMA) schemes will not be sufficient  and also pure non-orthogonal multiple access (NOMA) methods will not have the flexibility to support wireless connectivity for devices with different service requirements \cite{NOMA2006}. Therefore, new multiple access/resource allocation and interference management methods will need to be developed for these cell-less networks given the limited spectrum resources. In the following section, we propose a new method for massive multiple access in such a network that utilizes the cell-less 6G network architecture to support massive wireless connectivity.

\section{Delta-Orthogonal Multiple Access (D-OMA)}
This section first briefly discusses the main principle of NOMA compared to that of OMA. It then discusses the potential application of massive in-band NOMA in the new cell-less network architecture. Finally, the new D-OMA scheme is discussed and evaluated.

\subsection{OMA Versus NOMA}
\begin{figure}[!htb]
		\centering		\includegraphics[height=5.8cm, width=8.7cm]{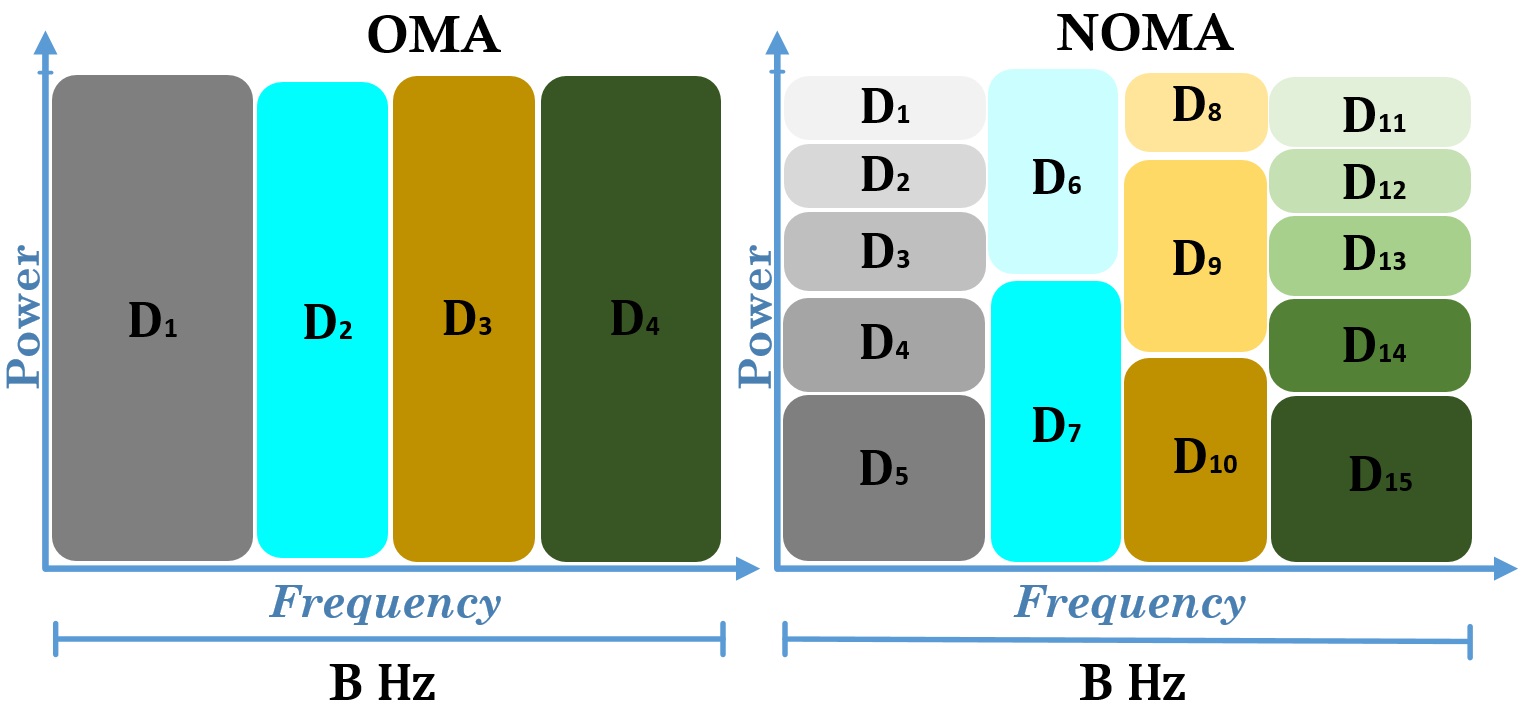}
	\caption{The concept of NOMA to serve multiple wireless devices within the same sub-band. }\label{OMANOMA}
	\end{figure}

OMA has been used for cellular generations 1G through 4G.  Due to the orthogonality among different sub-carriers and the relatively high bandwidth separation requirements among them, orthogonal frequency-division multiple access (OFDMA),  which is used on 4G networks, may not provide an efficient solution for future generation networks. Therefore, the NOMA  technique has been adopted lately by the 3GPP release-16 standards (5G) \cite{NOMA5G}.  
Generally, NOMA utilizes the concept of  superimposing many signals at the power domain within the same sub-band and using successive interference cancellation (SIC) at the receiver's end to filter out the undesired interfering signals. Using NOMA, every single OMA sub-band can serve multiple devices simultaneously and in this process a higher portion of transmission power is given to those with lower link quality (Fig. \ref{OMANOMA}).

Specifically, in an $M$ device/user NOMA cluster, for downlink transmission, the AP will send $x=\sum_{m=1}^M \sqrt{P_m}s_m$ such that $\sum_{m=1}^MP_m\leq P_t$, where $P_m$ is the transmission power allocated to the $m$-th NOMA device, $s_m$ is the signal to be send to the $m$-th device and $P_t$ is the maximum power budget assigned to a particular NOMA cluster's sub-band. The received signal at the $m$-th device is then given by $y_m=h_mx+w_m$, where $h_m$ is the complex channel gain between the AP and the $m$-th device, $w_m$ is the additive white Gaussian noise (AWGN) plus other clusters' interference signal. 
If channel gains of devices within a certain cluster are ordered as  $h_1\leq \dots \leq h_M$, then the transmission power levels will be assigned to every device  such that $P_1\geq \dots \geq P_M$.  At the receiver side, the interfering signals from devices with higher received powers are removed through SIC operation until the desired signal is decoded.  Accordingly, the achievable rate at the $m$-th device within a certain NOMA cluster of size $M$ is given by
\begin{equation}
    R_m=B \log_2 
    \left(
    1+
    \frac{P_m\gamma_m}{\sum_{j=m+1}^M P_j\gamma_m+1}
    \right),
\end{equation}
where $\gamma_m=\frac{|h_m|^2}{I_m+N_m}$ in which $I_m$ and $N_m$ represent the inter-cluster interference (ICI) and the AWGN powers at the input of the $m$-th device, respectively.
Generally, every sub-band will serve a single NOMA cluster. The devices within a certain cluster will suffer from two types of interferers, namely, intra-NOMA interference (INI) caused by residual unfiltered NOMA interference signal that is caused by other NOMA devices within the same cluster and the inter-cluster interference (ICI) that is caused by using the same sub-band by other nearby clusters. 
The size of a NOMA cluster can be considered as a design parameter to achieve trade-off among several factors, namely, the data rate requirements of the devices/users, the complexity level at NOMA receivers, the overall power budget per NOMA cluster, and the NOMA device immunity to INI, ICI, and SIC-based error propagation. 

\subsection{Massive In-Band NOMA}

Under the new 6G cell-less architecture, all nearby APs will be able to jointly cooperate with each other to serve a certain NOMA cluster simultaneously.
Cooperation among nearby APs will be generalized in a sense that all devices within their coverage area will be served by all APs  (either through reception of multiple copies of their desired signal from different APs or by mitigation of interference which is caused by adjacent non-serving APs). Such a diversity enhancing technique will add more degree of freedom in increasing NOMA cluster size.  This is referred to as the {\em massive in-band NOMA} scheme. Generally, under full APs connectivity, if a set of $K$ APs are used to serve a single $M$-device NOMA cluster throughout a certain sub-band of bandwidth $B$, the downlink rate at the $m$-th device within that cluster can be given as

\begin{equation}
    R_m=B \log_2
    \left( 
    1+\frac{\sum_{k=1}^KP_{m,k}\gamma_{m,k}}{\sum_{k=1}^K\left(
    \sum_{j=\beta+1}^MP_{j,k}
    \right)\gamma_{m,k}+1}
    \right),\label{mNOMAEq}
\end{equation}
where $\gamma_{m,k}=\frac{|h_{m,k}|^2}{N_m}$ and $\beta$ is the global ordering rank of the $m$-th device with respect to all serving APs.
The value of $\beta$ depends on the unified link quality metric used in ordering the devices of a certain cluster with respect to all serving APs.
Note that $I_m$ here is set to zero due to the comprehensive cooperation among APs and orthogonality among NOMA clusters. 

\begin{figure}[!htb]
		\centering		\includegraphics[height=7.5cm, width=8.9cm]{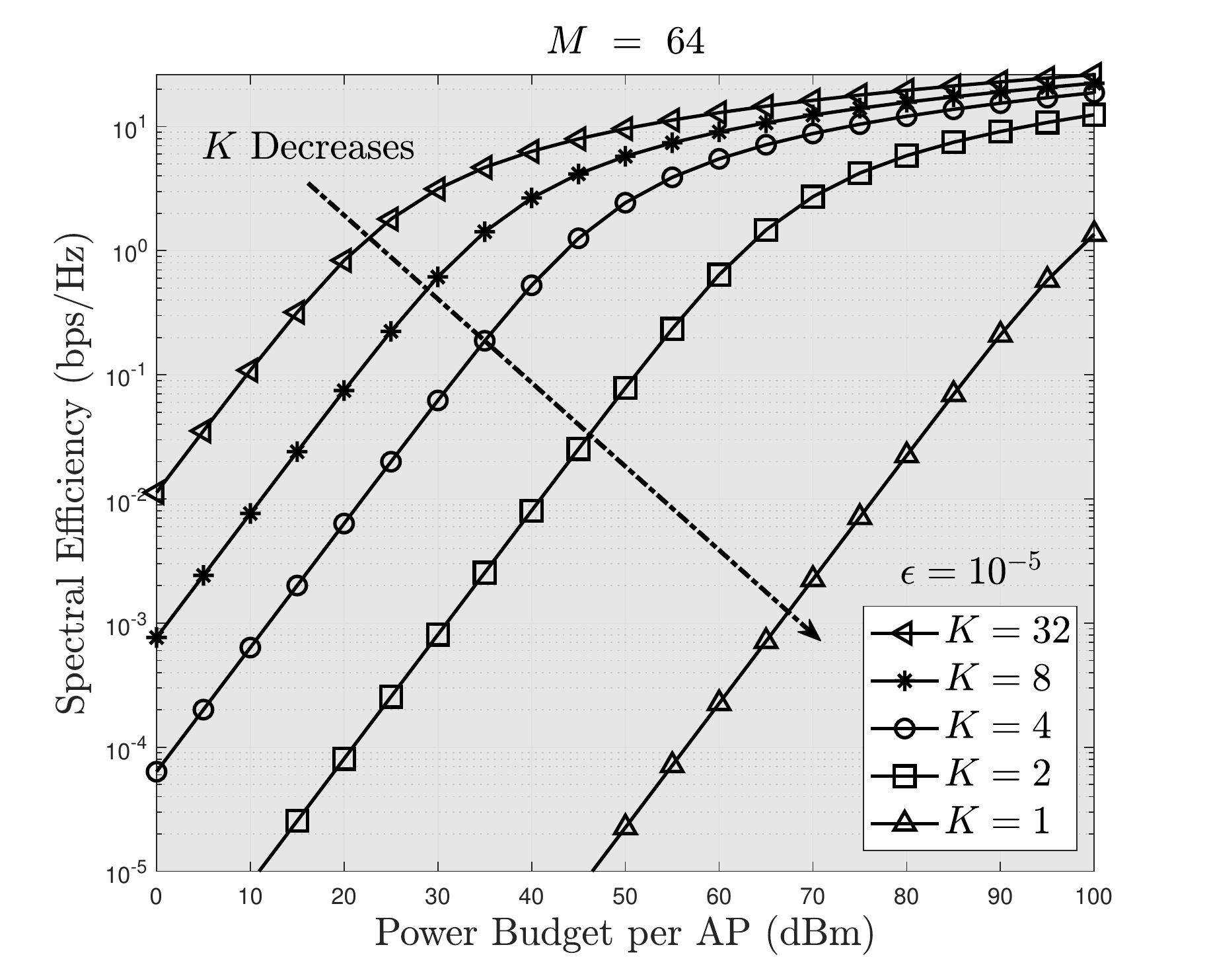}		\caption{ Example of massive in-band NOMA performance.}\label{mNOMA}
	\end{figure}

Fig. \ref{mNOMA} shows how does the new 6G architecture enhance the performance of an arbitrary NOMA device operating under massive in-band NOMA scheme. Note that in such a scheme the large INI component caused by massive NOMA cluster size is compensated by increasing the number of serving APs per NOMA cluster.

In Fig. \ref{mNOMA}, the asymptotic spectral efficiency of the so called {\em ($\epsilon$-outage)-capacity} is plotted,
where $\epsilon$ is the maximum allowable probability of outage per NOMA device to achieve a certain spectral efficiency.
Interestingly, the spectral efficiency performance of massive in-band NOMA scheme enhances significantly as the number of cooperating APs increases (diversity gain can be noticed by the significant increase of the slopes of performance curves as $K$ increases).
However, a very large cluster size in such a scheme will result in high complexity at NOMA receivers which may not be suitable for mMTC devices with simple  design and minimal power consumption requirements. Nevertheless, massive in-band NOMA may be enabled for some services where the size and complexity of the wireless devices and the required amount of power consumption are feasible.
\subsection{D-OMA: A New Multiple Access Scheme for 6G}
\begin{figure}[!htb]
		\centering		\includegraphics[height=7.5cm, width=8.25cm]{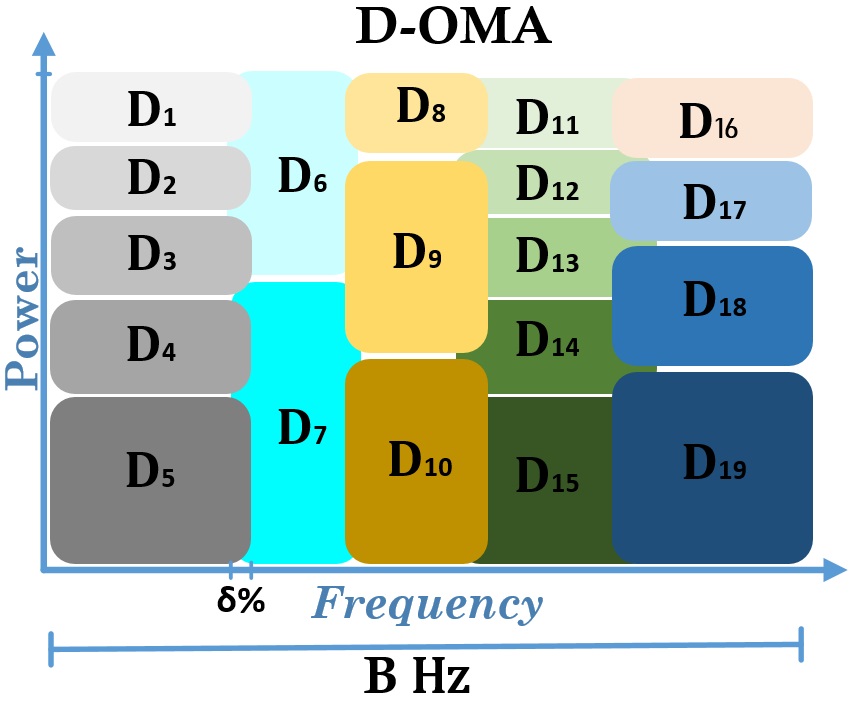}
	\caption{An example of the proposed D-OMA scheme.}\label{DOMA}
	\end{figure}

As has been stated above, the high complexity and increased power consumption at terminal devices are two major obstacles that hamper the deployment of in-band NOMA clusters at the massive scale regime. 
Here, we propose a new massive multiple access scheme, namely delta-orthogonal multiple access (D-OMA) that tackles these two problems and therefore would be suitable for 6G network architecture  and requirements (Fig. \ref{DOMA}). 

In D-OMA, we allow different NOMA clusters with adjacent frequency bands to overlap by an amount of $\delta$ percent of their maximum allocated sub-band (i.e. $\delta = 0$ corresponds to traditional power-domain NOMA).
In order to compensate for the ICI introduced by overlapping NOMA clusters, the size of NOMA cluster may be decreased such that the overall sum of INI and ICI remains almost same as the value of the INI when non-overlapping massive in-band NOMA clustering is used.
Accordingly, the spectral efficiency achieved by massive in-band NOMA cluster is maintained by adding more clusters within the same allocated overall spectrum ($B$ Hz in Figs. \ref{OMANOMA} and \ref{DOMA}). At the same time, by decreasing the sizes of different NOMA clusters, the level of complexity requirements and power consumption on different NOMA terminal devices will be significantly decreased while maintaining the same performance requirements as before.  

Generally, the percentage of overlapping among NOMA clusters ($\delta$) is considered as a design parameter in the global optimization problem that achieves optimal clustering for NOMA devices as well as their power allocation.  Under the proposed access scheme, the transmission rate at the $m$-th device within a certain NOMA cluster is given by
\begin{equation}
    R_m=B \log_2
    \left( 
    1+\frac{\sum_{k=1}^KP_{m,k}|h_{m,k}|^2}{\sum_{k=1}^K\Lambda_k|h_{m,k}|^2+\delta I_{\text{ICI}}+N_m}
    \right),
\end{equation}
where $\Lambda_k=\sum_{j=\beta+1}^{\mathcal{M}} P_{j,k}$ with $\mathcal{M}\leq M$ being the new cluster size compared to Eq. (\ref{mNOMAEq}).
Note that for the same $K$, a smaller $\mathcal{M}$ will result in a smaller $\Lambda_k$, and hence, a higher signal-to-interference-plus-noise ratio (SINR) at the input of the $m$-th device. 

\begin{figure}[!htb]
		\centering	\includegraphics[height=7.5cm, width=8.9cm]{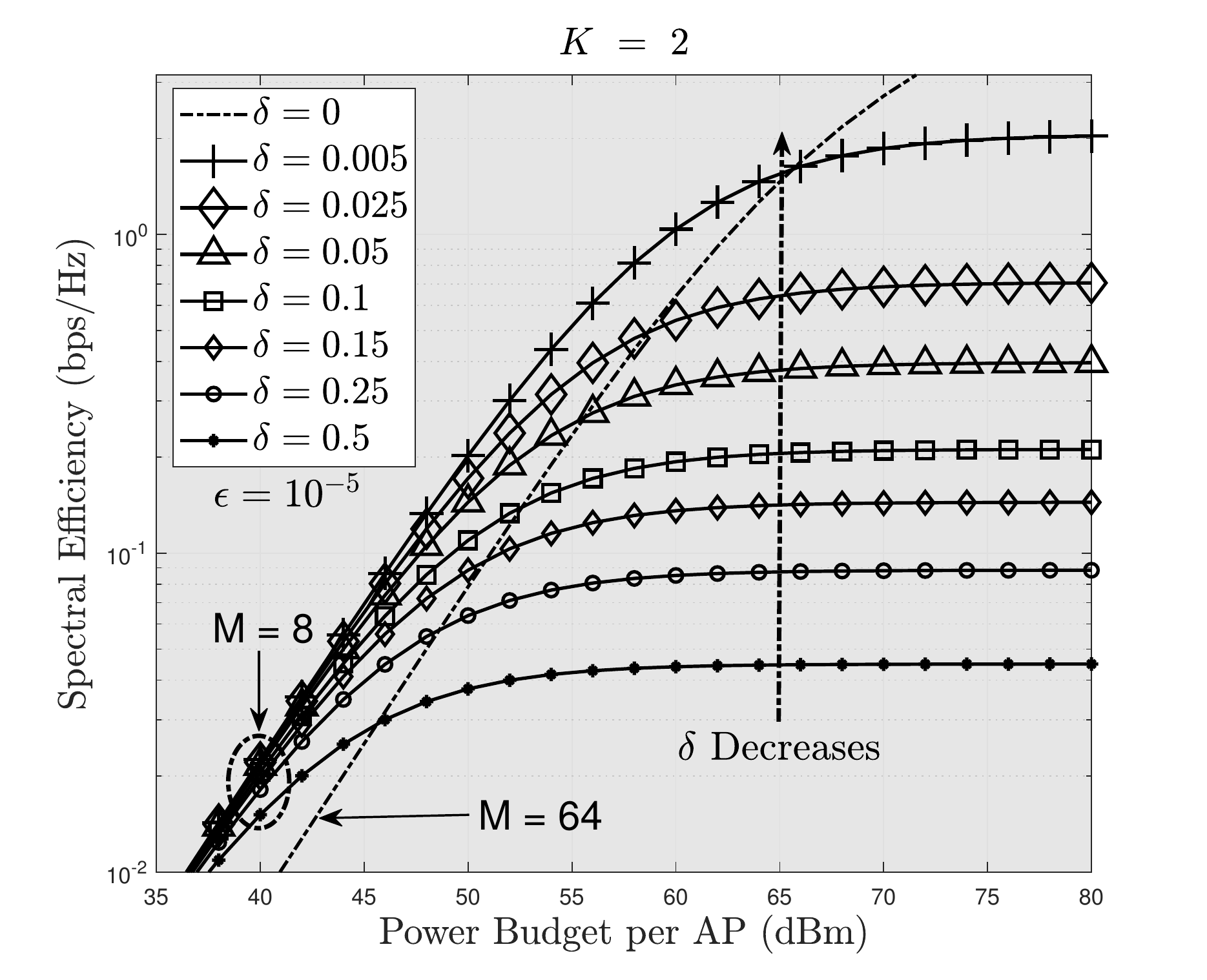}		\caption{An example of D-OMA performance.}\label{Delta_DOMA}
	\end{figure}

Fig. \ref{Delta_DOMA} illustrates the performance of the proposed access scheme compared to the massive in-band NOMA scheme presented in previous section. 
In this figure, the asymptotic {\em ($\epsilon$-outage)-capacity} per device is plotted with two different cluster sizes and different values of overlapping percentage $\delta$. 
We consider $I_{\text{ICI}}$ to be equal to $10\%$ of the maximum power budget per AP and set $N_m=1$.
The number of devices served by a single sub-band ($B$ Hz) is assumed to be $M = 8$ and $M = 64$. The overall sum-rate achieved by the cluster with larger number of devices (e.g. $M=64$) is greater than that with lower $M$ (e.g. $M=8$). For $M=64$, in Fig. \ref{Delta_DOMA}, the value of $\delta$ is assumed to be 0.  The proposed scheme may be applied to both uplink and downlink transmissions.  However, applying D-OMA at the uplink may be more difficult due to the increased requirements of control signal exchange between cooperating APs and terminal devices.
The performance of D-OMA compared to that of massive in-band NOMA will depend on the cluster size, the amount of overlapping in the spectrum shared by the clusters.
With D-OMA, the cluster size (and hence the complexity of SIC) can be reduced significantly when compared to massive in-band NOMA with large cluster sizes while providing similar performance.  
 Note that, performance of D-OMA in terms of area spectral efficiency as well as energy efficiency can be obtained by conducting a system-level simulation considering devices' spatial distributions and wireless channel propagation characteristics. Nevertheless, the insights extracted from the studied example will be still valid.

\begin{figure}
    \centering
   \includegraphics[height=5.6cm, width=3.5in]{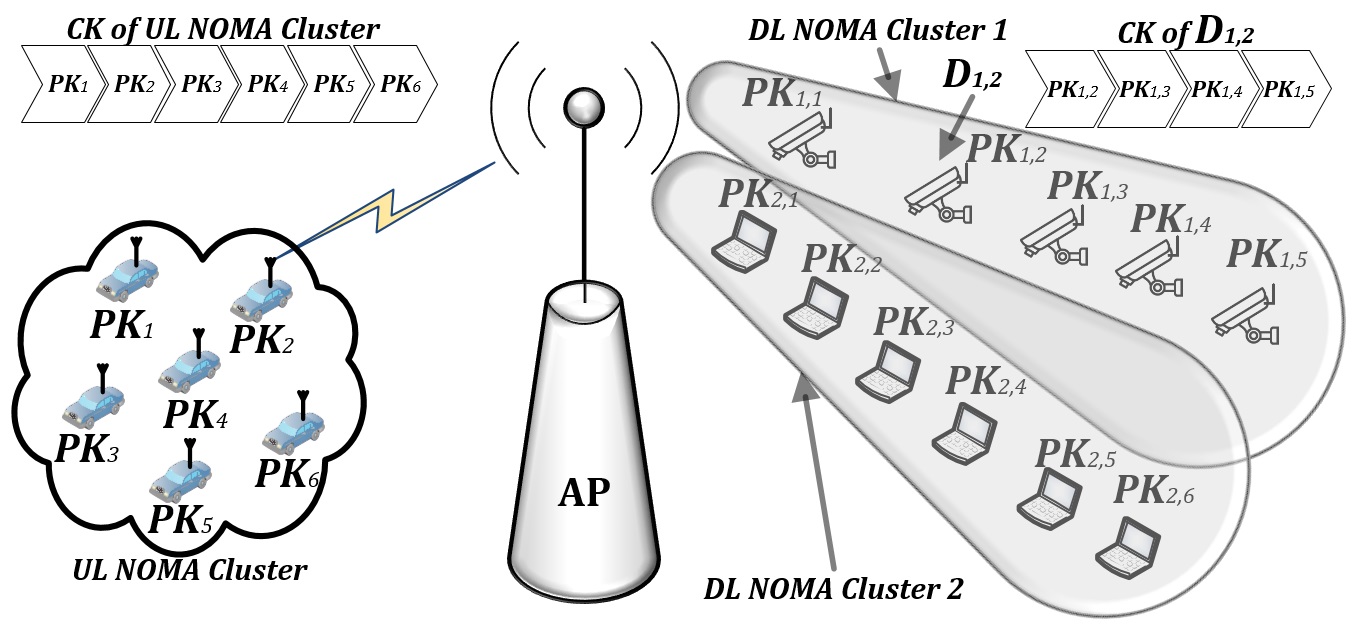}
    \caption{\textsf{Provisioning of security with D-OMA-based wireless access.}}
   \label{Dist_Encrypt}
\end{figure}

An important aspect of D-OMA scheme will be provisioning of security in terms of very low intercept probability and significantly enhanced higher level of encryption. This can be achieved based on the idea of 'Jigsaw Puzzle' in which a group of oddly shaped interlocking and tessellating pieces are combined together to produce a final usable image (Fig. \ref{Dist_Encrypt}). 

In the uplink, every NOMA device will have a certain partial key (PK) that represents a part of a final cluster key (CK). At the receiver end, signals from different devices within the same cluster will be decoded using SIC (at the code-word level) and then different PKs will be assembled to form a final CK. Generally, the CK is considered as the decrypt key of received data from all devices within that cluster and data of every device will not be reconstructed without full acknowledgement of that key. That is,  an eavesdropper will need to decode data from all users at the same time and same order which will be very difficult when a huge number of terminal devices are transmitting to a single uplink receiver.
This is due to the fact that only the receiving APs will be aware of the required decoding ordering during SIC operation and of the content of every NOMA cluster as well (when different clusters are  partially overlapping).
Additionally, NOMA cluster ordering represents a unique function of links' quality between cluster elements and the set of cooperating APs which in turn a function of the locations of the APs and the cluster devices as well.
Hence, an eavesdropper will receive signals from different devices at a different random weights (without knowledge of the original cluster members' ordering) which makes it extremely difficult (if not impossible) to decode and then extract the CK of a certain cluster.

In the downlink, the $m$-th NOMA device will have the ability to decode a set of $m-1$ signals from other devices within the same cluster ($m=1, \dots, M$). 
Accordingly, it is practical that a final CK at certain downlink NOMA device is combined from the set of PKs of NOMA devices with lower chanel quality than that device (see Fig. \ref{Dist_Encrypt}). 
As a result, the device with the highest link quality (with resepect to all serving APs) will have the best security while the device with the worst link quality will utilize a single-PK protection (which is identical to conventional wireless link security such as WEP, WPA and WPA2 standards \cite{Security7467419}). 
For example, if devices in cluster 1 in Fig. \ref{Dist_Encrypt} are numbered based on their overall SINR, to decrypt the desired data at device $D_{1,2}$, the required key will be the combined PKs of devices $D_{1,2}, D_{1,3}, D_{1,4},$ and $D_{1,5}$.  
With this set up for downlink security through distributed encryption, any eavesdropper attempting to hack into the $m$-th NOMA device must be able to decode $D_{1,2}, D_{1,3}, D_{1,4},$ and $D_{1,5}$ which requires having the appropriate SIC decoding order, and such a scenario may happen with extremely low probability.
Accordingly, the proposed new D-OMA-based wireless security concept will decrease the intercept probability and enhance the encryption strength level of the wireless system significantly. 
    
 \section{Implementation Issues and Open Research Challenges }
In this section, practical implementation issues of D-NOMA  are discussed.
 
 \begin{enumerate}[\huge .]
     \item {\em Power allocation for D-OMA devices/users}: The idea of D-OMA is motivated from the  assumption that comprehensive cooperation among nearby APs is utilized.
     Accordingly, transmission power allocation methods used in conventional NOMA schemes cannot be applied to D-OMA. Note that many researches were conducted in the literature to optimize NOMA transmission power for single-cluster single-antenna \cite{Shipon7557079}, multiple clusters multiple-antennas \cite{ShiponMIMO}, and CoMP scheme \cite{Shipon2018}.
    In the proposed cell-less network architecture, the coordinating APs must jointly allocate power fractions for devices within a certain cluster such that the sum rate is maximized while a certain minimum rate per D-OMA device is maintained. 
     Such an optimization problem has to be solved efficiently and quickly since optimization parameters will change at the time-slot level in a dynamic environment. This eliminates the possibility of using tedious optimization algorithms such as exhaustive search.
     
     \item {\em Clustering of devices/users D-OMA}: The performance of D-OMA strongly depends on the selection of cluster size, cluster members, cluster overlapping percentage ($\delta$), and the set of cooperating APs.
     Generally, optimal selection of these parameters is very complicated (if not impossible), especially for ultra-dense networks with dynamic propagation environments.
     This may be solved by using some sub-optimal algorithms such as heuristic algorithms or deep reinforcement learning algorithms so that near-optimal clustering decisions may be taken without solving the global optimization problems \cite{8387430,8542687,8422168}. Also, in presence of device/user mobility and dynamic variation of the propagation environment, the spatio-temporal correlation of interference \cite{Spatio7976286} experienced by devices/users in a cluster will vary dynamically. This variation in interference correlation can be exploited in clustering the devices/users as well as allocating power in order to enhance system performance. 
     
     \item {\em Efficient SIC method for D-OMA}: In conventional NOMA, devices are ordered from one AP's prospective and SIC is achieved according to that ordering.
    The same SIC technique cannot be used for D-OMA since different devices may have different ordering from the perspective of multiple cooperating APs. 
     This will pose a new challenge in the design of SIC methods at NOMA receivers.
     To decrease the complexity of a D-OMA receiver, some sub-optimal low-complexity methods that assign the same ordering for every device with respect to all serving APs will be required. 
     This can be achieved by selecting the best ordering for D-OMA cluster members with respect to all serving APs/BSs.
     
     \item {\em Spectrum variations}:
     New wireless generations including 5G and 6G will adopt a very large range of frequencies that in turn will have variable system requirements in terms of transmission power and network structures.
     Accordingly, the design of D-OMA parameters including the overlapping percentage ($\delta$) must take the used frequency range into consideration.
     Specifically, for high frequency bands the severe path-loss will be compensated by more transmission power and focused beams. 
     This may cause very high ICI impact if $\delta$ is not selected properly.
     On the other hand, lower frequency bands will suffer from less propagation losses and hence will have less transmission power requirements. 
     Accordingly, for larger values of $\delta$, the performance of D-OMA is likely to be better in lower frequency bands than that with higher frequency bands. 
     
     \end{enumerate}
 	
\section{Conclusion}
We have proposed a new multiple access scheme for massive wireless connectivity in B5G/6G wireless networks. This scheme exploits the idea of coordinated multipoint transmissions from a large number of access points (APs)/base stations (BSs) and non-orthogonal multiple access (NOMA) among users in a spectrum band along with partial overlapping of spectrum bands among  NOMA clusters.  The system parameters such as cluster size, amount ($\%$) of spectrum overlap among clusters, number of cooperating APs/BSs can be chosen to achieve a desired tradeoff between performance and complexity (e.g. due to SIC decoding). This scheme can provide significant spectrum efficiency gain that will be required for 6G systems. Implementation issues  and related challenges for this proposed scheme have been outlined. 
	
 \bibliographystyle{IEEEtran}
\bibliography{IEEEabrv,yasser}
\end{document}